\documentclass{optica-article}

\journal{opticajournal} 

\articletype{Research Article}

\usepackage{lineno}
\usepackage{blkarray} 
\newcommand{\matindex}[1]{\mbox{\scriptsize#1}}
\usepackage{isomath}
\usepackage{comment}

\begin{document}

\title{Direct Polarization-Entangled Photon Pair Generation Using Domain-Engineered Nonlinear Crystals}

\author{Anatoly Shukhin,\authormark{1,$\dagger$,*} Inbar Hurvitz,\authormark{2,$\dagger$} Leonid Vidro,\authormark{1} Ady Arie,\authormark{2} and Hagai S. Eisenberg\authormark{1}}

\address{\authormark{1}Racah Institute of Physics, Hebrew University of Jerusalem, Jerusalem 91904, Israel\\
\authormark{2}School of Electrical and Computer Engineering, Fleischman Faculty of Engineering, Tel Aviv University, Tel Aviv 69978, Israel\\
\authormark{$\dagger$}These authors contributed equally to this work.}

\email{\authormark{*}anatoly.shukhin@mail.huji.ac.il} 


\begin{abstract*} 
A bi-photon polarization-frequency entanglement source was realized by shaping the phase-matching function of a poled KTP crystal. It provides a simple method to achieve either polarization or spectral entanglement in a simple collinear setup, based on single-pass SPDC and a dichroic (or polarizing) beam splitter. This is a robust and cost-effective configuration that can be easily implemented outside the laboratory environment. We characterized the source by two approaches: reconstructing the density matrix of the generated state with quantum state tomography, by recording the coincidences across 16 mutual polarization settings, in addition to a new method based on quantifying the symmetry of the joint spectral intensity by swapping between the signal and idler wavelengths. The polarization-entangled source violates the Clauser-Horne-Shimony-Holt inequality with a measured $S=2.747\pm 0.004$. We also measured the polarization entanglement visibility in two mutually unbiased polarization bases and evaluated the squeezing level by characterizing the reduction in visibility at high pump power. 
\end{abstract*}

\section{Introduction}

Polarization-entangled photon pairs are a fundamental resource for many quantum technologies and experiments in quantum optics. Their applications range from quantum teleportation\cite{bouwmeester1997experimental}, entanglement swapping\cite{EntanglementSwapping1998}, and entanglement-based quantum cryptography \cite{Ekert} to fundamental tests of quantum mechanics and the exploration of the nature of physical reality \cite{giustina2017significant,zeilinger1999experiment}.

The most commonly used approach for generating these states is by leveraging nonlinear optical processes, specifically spontaneous parametric down-conversion (SPDC) and spontaneous four-wave mixing (SFWM) in bulk nonlinear crystals and nonlinear optical waveguides. Due to their tunability, and their generation of bandwidth-limited wave-packets at room temperature, these processes are fundamental to modern quantum technologies \cite{anwar2021entangled,eisaman2011invited}. In addition to polarization, SPDC and SFWM can generate photon pairs entangled in other degrees of freedom, including frequency-time \cite{gianani2020measuring,strekalov1996postselection}, position-momentum \cite{chan2007transverse,just2013transverse,DiDomenico:22}, and orbital angular momentum \cite{yesharim2023,Osorio2009}, as well as the creation of hyper-entangled states \cite{Fedrizzi_hyper}. SPDC, which is proportional to a second-order susceptibility $\upchi^{(2)}$, generally exhibits higher efficiency compared to SFWM \cite{samantaray2024frequency}, which relies on third-order non-linear susceptibility $\upchi^{(3)}$. Thus, our focus in this work is on the SPDC process.

Several well-established schemes exist for the efficient and high-fidelity generation of photonic polarization entangled states. These include non-collinear schemes \cite{Kwiat_1995}, as well as collinear: post-selection at a beam splitter\cite{kuklewicz2004high}, a Sagnac-like loop \cite{kim2019pulsed,weston2016efficient,li2015cw,jin2014pulsed}, crossed crystals \cite{pelton2004bright,hubel2007high}, the "folded sandwich"
\cite{steinlechner2013phase,steinlechner2015sources}, and more \cite{anwar2021entangled}. However, each of these schemes has its drawbacks. For instance, post-selection at a beam splitter halves the rate, while schemes like the Sagnac loop and crossed crystals do not fully utilize the pump power. Certain methods, such as the "folded sandwich", are not intrinsically stable and require active stabilization \cite{anwar2021entangled}. Furthermore, these schemes typically suffer from complexity in construction and alignment, lack of compactness, and insufficient temperature and mechanical stability, which are crucial for scaling up in real-world applications.

Recently, it was demonstrated that polarization-entangled photon pairs can also be obtained in a single-pass configuration. Different methods were demonstrated by using modal birefringence in a microcavity \cite{Francesconi:23}, the positive and negative orders of quasi phase-matching \cite{Laudenbach2017}, and by employing domain-engineered bulk nonlinear crystals \cite{kuo2020demonstration,Kaneda:19}. A common feature of these schemes is that the nonlinear medium inherently supports two type-II collinear non-degenerate SPDC processes: $|H_{\omega_1}V_{\omega_2}\rangle$ and $|V_{\omega_1}H_{\omega_2}\rangle$, thereby forming a polarization entangled state. However, all of these demonstrations do not shape the two processes spectrally to achieve round lobes in the joint spectral intensity (JSI). This spectral shaping is crucial for achieving entanglement between two single-modes and thus, a source of high fidelity entanglement that is suitable for the entanglement swapping protocol\cite{humble2008effects}.

In this work, we present a single-pass single-crystal post-selection-free source of single-mode polarization entangled states at telecom wavelengths, based on SPDC in a domain-engineered bulk $\mathrm{KTiOPO_4}$ (KTP) crystal. Our source offers several key advantages:
the scheme features a simple and compact collinear design, requiring only a few optical components; the presented scheme eliminates the need for interferometers such as Sagnac, Michelson, or Mach-Zehnder, resulting in improved mechanical stability without the need for active stabilization; the source can perform well under both pulsed and continuous pumps. We demonstrate that for a femtosecond pulsed pump, the crystal can be designed to produce a round JSI of each process $|H_{\omega_1}V_{\omega_2}\rangle$ and $|V_{\omega_1}H_{\omega_2}\rangle$, making this scheme suitable for quantum repeaters requiring high-visibility interference between two independent sources and therefore pure heralded single photons from each source. 

\textcolor{black}{A key aspect of our work is directly extracting the two-photon polarization density matrix from the JSI measurement. This approach enables a fully spectral-based characterization of the polarization entanglement, providing a powerful alternative to traditional quantum state tomography.}
Our work presents a promising advancement in generating polarization entangled states, offering simplicity, efficiency, and versatility for many quantum applications.

\section{SPDC in domain-engineered crystals}
SPDC is a process of interaction between three modes of electromagnetic field in a nonlinear crystal, in which the annihilation of one photon from one mode (pump) is accompanied by the simultaneous creation of two photons into two other modes, traditionally referred to as the signal and idler. For this process to be efficient, the amplitudes of photon pair creation at different points in time and space (crystal volume) must constructively interfere. This leads to the phase-matching conditions that connect the frequencies and wave vectors of the three interacting fields: 

\begin{equation}
    \omega_{p} = \omega_{s} + \omega_{i},
\end{equation}
\begin{equation}\label{k_phase_matching}
    k_{p}(\omega_p) = k_{s}(\omega_s) + k_{i}(\omega_i),
\end{equation}
where $\omega_{l}$ and $k_{l}~(l = p,s,i)$ are the frequencies and wave numbers of the pump, signal and idler fields, respectively. Here, $k_{l} = n(\omega_l)\omega_{l}/c$, where $n(\omega_l)$ is the refractive index of the crystal at the frequency $\omega_{l}$, and $c$ is the speed of light. In this work, we focus on collinear type-II SPDC. Due to the collinearity of the process, Eq. \ref{k_phase_matching} is scalar, with all the wave vectors pointing along the crystallographic X-axis of the nonlinear crystal.

The state vector describing the generated field can be calculated in the first order of the perturbation theory and is expressed as \cite{keller1997theory}:
\begin{equation}\label{eq:PDC_State}
\begin{split}
    |\psi\rangle =& |0\rangle 
    + \zeta\iint  d\omega_{s}d\omega_{i}f(\omega_{s},\omega_{i}) a^{\dagger}_H(\omega_{s}) a^{\dagger}_V(\omega_{i}) |0\rangle,
\end{split}
\end{equation}
where $|0\rangle$ denotes the vacuum state, and $a^{\dagger}_H(\omega_{s})$ and $a^{\dagger}_V(\omega_{i})$ represent the creation operators for the signal and idler modes. The function$f(\omega_{s},\omega_{i})$ is the joint spectral amplitude (JSA) of the two-photon field, which describes the amplitude of generating the signal and idler photons at the respective frequencies $\omega_{s}$ and $\omega_{i}$. The square of the absolute value of the JSA is the JSI. Aside from the central frequencies and spectral bandwidths of the signal and idler photons, the JSA also reveals spectral correlations inherent to photon pairs (or the absence thereof, which is desired for generating pure heralded single photons \cite{zielnicki2018joint}). As will be shown in the following sections, with proper design of the crystal's nonlinear susceptibility $\upchi^{(2)}$ -- and hence, proper shaping of the JSA -- these spectral correlations can lead to polarization entanglement between the two photons generated within the crystal. This allows for forming an entangled state using only a domain-engineered crystal without additional lossy manipulation of the generated photons, as in other methods.

In the low-gain regime, when the normalized coupling coefficient $\upkappa = \upchi^{(2)}E_p L\omega/cn(\omega) < 1$ (where $E_p$ is the pump field amplitude, $L$ is the crystal length, and $n(\omega)$ is the frequency-dependent refractive index of the crystal) \cite{hurvitz2023,Yesharim_review_2025}, the JSA can be expressed as the product of the spectral amplitude of the pump field, $P(\omega_s,\omega_i)$, and the phase-matching function of the crystal (PMF), $\upPhi(\upDelta k(\omega_s,\omega_i))$:
\begin{equation}
f(\omega_{s},\omega_{i}) = P(\omega_s,\omega_i) \cdot \upPhi(\upDelta k(\omega_s,\omega_i)).
\end{equation}
The PMF in the frequency domain is obtained via the Fourier transform of the crystal's poling pattern \cite{kaneda2021generation}, which allows for achieving the desired amplitude and phase of the PMF (including analytical solutions), by shaping the crystal's nonlinear susceptibility\cite{shiloh2012spectral,leshem2014experimental}.
Using this technique, the crystal employed in this work is engineered so that its PMF in the frequency domain consists of two lobes with a $\pi$ phase difference between them (Fig. \ref{fig:JSI}(a),(b)). When pumped at a wavelength of 780 nm, the central wavelengths of the lobes were designed to be at 1548 nm and 1572 nm. The spectral bandwidth (FWHM) of each lobe is 13 nm.
The crystal is 4 mm long and has a poling period of $46~\mu m$. The poling pattern of the crystal results in the following phase-matching function:
\begin{equation}
\upPhi(\upDelta k(\omega_s,\omega_i))=\frac{1}{\sqrt{2\pi\sigma}} \left( \exp\left( -\frac{(\upDelta k - a)^2}{2\sigma^2} \right) - \exp\left( -\frac{(\upDelta k + a)^2}{2\sigma^2} \right) \right),
\end{equation}
where $\sigma = 333~m^{-1}$ and $a = 2700~m^{-1}$.

The state generated at the output of the crystal can be written  as follows:
\begin{equation} \label{eq:initial state}
    |\psi\rangle =\frac{1}{\sqrt{2}} \Big(|\omega_{1}\rangle|H\rangle \otimes |\omega_{2}\rangle|V\rangle-|\omega_{1}\rangle|V\rangle \otimes|\omega_{2}\rangle|H\rangle \Big),
\end{equation}
where $\omega_{1}$ and $\omega_{2}$ represent the lower and higher central frequencies of the JSA lobes, the minus sign reflects the $\pi$-phase between the PMF (and JSA) lobes, and the $1/\sqrt{2}$ factor accounts for the equal pair rate generated for each lobe.
This state can be converted to a polarization-entangled state by separating the bi-photon field in a fiber add-drop filter (ADF). Alternatively, a free-space dichroic mirror with the cut-on frequency $\omega_{0}$ between the two lobes can split the two created photons into separate paths. After such a transformation, which traces out the frequency degree of freedom, the state becomes:
\begin{equation}\label{eq:state after dichroic}
    |\psi_{ADF}\rangle =\frac{1}{\sqrt{2}} \Big(|H_{\omega_{1}}V_{\omega_{2}}\rangle - |V_{\omega_{1}}H_{\omega_{2}}\rangle\Big).
\end{equation}
Alternatively, if the photon pairs in the original state are separated by a polarizing beam splitter (PBS), the polarization degree of freedom is traced out, forming a frequency-entangled state:
\begin{equation}\label{eq:state after PBS}
    |\psi_{PBS}\rangle =\frac{1}{\sqrt{2}} \Big(|\omega_{1_{H}}\omega_{2_{V}}\rangle-|\omega_{2_{H}}\omega_{1_{V}}\rangle\Big).
\end{equation}

This paper focuses on polarization-entanglement, whereas the realization of frequency entanglement with shaped crystals can be found in Ref. \cite{Shukhin2024,morrison2022frequency}.


\section{Experimental results}\label{experiment}

\subsection{Generation of Polarization-Entangled Photon Pairs}

\begin{figure}[t]
  \centering
  \includegraphics[width=1\textwidth]{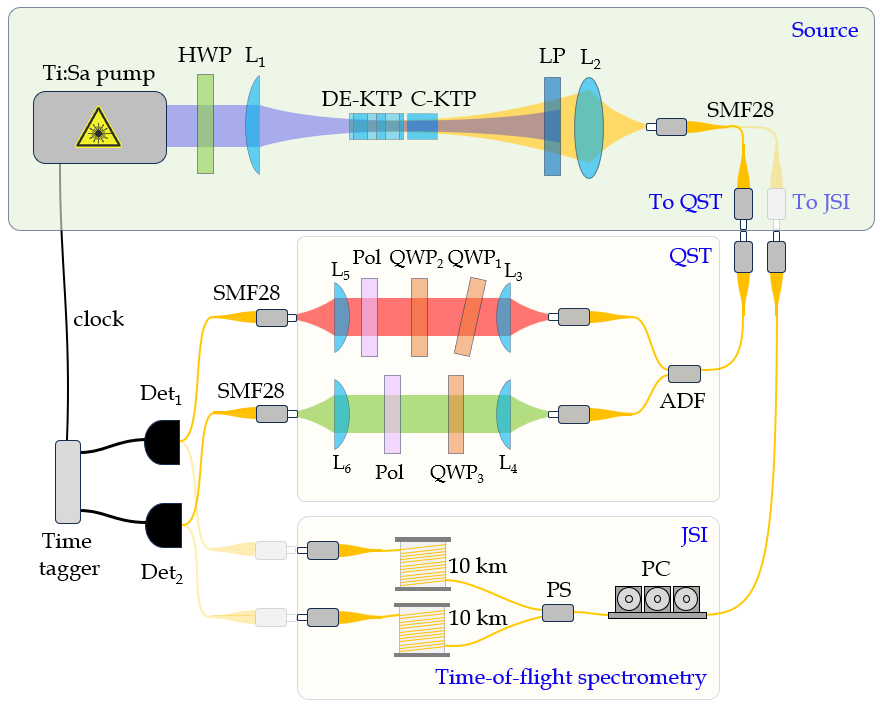}
\caption{Experimental scheme for direct generation and characterization of polarization-entangled photon pairs via SPDC in domain-engineered KTP. Here Ti:Sa is the Titanium-sapphire femtosecond laser, HWP -- half-wave plate, L$_{1-6}$ -- lenses, DE-KTP -- domain-engineered KTP crystal, C-KTP -- temporal-walk-off compensation crystal, LP -- longpass filter, SMF28 -- single-mode optical fiber at 1560 nm, ADF -- fiber add-drop filter, QWP$_{1-3}$ -- quarter-wave plates, Pol -- linear polarizer, PC -- fiber polarization controller, PS -- fiber polarization sorter, Det$_{1,2}$ -- superconducting nanowire detector \cite{scontelSNSPD}. QST -- quantum state tomography part, JSI -- part of the setup intended for the JSI measurements based on the time-of-flight spectrometry.}
\label{fig:setup}
\end{figure}

The experimental setup is illustrated in Fig. \ref{fig:setup}. The generation of polarization-entangled photon pairs occurs in the top part.

A mode-locked Ti-Sapphire laser at a wavelength of 780 nm was used as the pump source. The laser spectral bandwidth was approximately 7\,nm, corresponding to a pulse duration of around 90\,fs for a bandwidth-limited sech$^{2}$-shaped pulse, which is typical for this type of lasers. The pulse repetition rate was 76\,MHz. A half-wave plate (HWP) was used to align the pump polarization with the crystallographic Y-axis of the nonlinear crystal, facilitating type-II SPDC via the corresponding d$_{24}$ nonlinear coefficient. The pump beam was then focused into the nonlinear crystal using lens L$_{1}$.

The SPDC medium was a domain-engineered KTP (DE KTP) crystal from \textit{Raicol Crystals}, with the poling pattern as discussed in the preceding section. The crystal was operated at room temperature ($\approx 23^{\circ}$C) without controlling its temperature.

An essential requirement for observing entanglement is the indistinguishability of the two corresponding amplitudes: $|HV\rangle$ and $|VH\rangle$. However, the birefringent delay between the H and V photons in the DE KTP, referred to as temporal walk-off, introduces temporal distinguishability. To compensate for this walk-off, an additional unpoled 2\,mm-long KTP crystal was placed after the DE KTP crystal, oriented perpendicularly in the Z-Y plane of the crystal. The difference in the walk-off between the $|H_{\omega_{1}}V_{\omega_{2}}\rangle$ and $|V_{\omega_{1}}H_{\omega_{2}}\rangle$ amplitudes was calculated to be 1.3\,fs, which is significantly smaller than the duration of the photon (617\,fs) and was therefore disregarded.

After exiting the crystal, the pump was suppressed using three long-pass dichroic mirrors with a cut-on wavelength of 950\,nm. Each filter has an attenuation of 99.5\% at the pump wavelength and 97.8\% transmission at 1560 nm. The collinearly generated biphoton field was then coupled, using lens (L$_{2}$), into a single-mode fiber (SMF-28). The fiber served as a spatial filter, ensuring the collected photons were spatially indistinguishable.


\subsection{Spectral Characterization}

In addition to the temporal indistinguishability discussed in the previous section, the spectral indistinguishability between the two terms is also crucial. The two terms in the state $|\Psi^{-}\rangle=(|H_{\omega_{1}}V_{\omega_{2}}\rangle-|V_{\omega_{1}}H_{\omega_{2}}\rangle)/\sqrt{2}$ correspond to the two lobes of the biphotons' JSA. Therefore, spectral indistinguishability requires that the  JSA of the generated photons be symmetric (or antisymmetric if the target is the singlet Bell state, which is the case in this paper) under swapping the frequencies of the signal and idler photons: \textcolor{black}{$f(\omega_{1},\omega_{2}) = e^{i\theta}f(\omega_{2},\omega_{1})$ with $\theta=\pi$ in our experiment}. Additionally, it is desired that each spectral mode remain pure after the frequency of the other mode is selected. This is not the case for elongated JSI lobes, where more than two spectral modes are entangled.

To verify these requirements, we measured the JSI of the generated photons using time-of-flight spectroscopy \cite{avenhaus2009fiber}, which leverages the natural capability of long and dispersive fibers to perform a frequency-time Fourier transform on incoming light.
Here, we assume that the JSA is a real-valued function, neglecting any joint/nonlinear spectral phase dependence and assuming that the linear spectral phase along the anti-diagonal axis has been eliminated through temporal walk-off compensation, retaining only the $\pi$ phase difference between the two lobes. As will be demonstrated in the following section, the validity of this assumption will be confirmed through quantum state tomography. This assumption, combined with the fact that the two lobes do not overlap when projected onto the frequency axes, allows us to replace the JSA with the square root of the experimentally measured JSI, where information about the spectral phase of the state is lost, enabling further symmetry evaluation. 

\begin{figure}[h]
  \centering
  \includegraphics[width=0.7\textwidth]{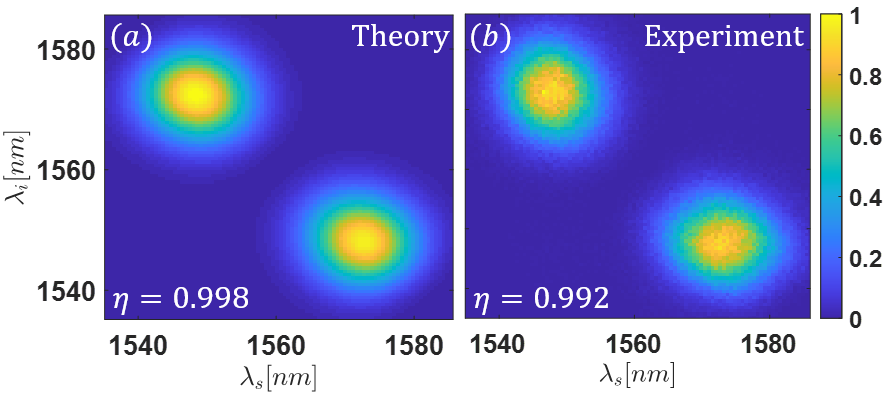}
\caption{Spectral characterization of the source. Simulated (a) and measured (b) JSIs, with overlap integral (Eq. \ref{eq:overlap_int}) values shown in each graph. The small difference between these values demonstrates good agreement between the experimental results and theoretical predictions.}
\label{fig:JSI}
\end{figure}

The spectral characterization setrup is presented in the JSI lower box of Fig. \ref{fig:setup}. The output fiber of the source was connected to the setup dedicated to JSI measurements, where a fiber polarization controller (PC) was followed by a fiber polarization sorter (PS) that splits the photons by polarization into two arms. In each arm, a 10\,km long SMF-28 optical fiber with a dispersion of 18\,ps/nm/km at 1560\,nm introduces a dispersive delay between the frequency components of the photons. Each 10\,km fiber was connected to a superconducting nanowire detector ($\mathrm{Det}_{1,2}$) \cite{scontelSNSPD}. The electrical signals from the detectors were recorded by a time-tagger to count coincidences. By measuring the number of coincidences between the two arms and a reference clock signal from the laser, we reconstructed the JSI (Fig. \ref{fig:JSI}(b)). The overall temporal resolution, determined by the jitter of our detection system, was measured to be approximately 150 ps. This value, along with the spectral delay after the 10\,km fibers (180\,ps/nm), corresponds to a spectral resolution of 0.83\,nm for the reconstructed JSI. For comparison, the calculated JSI based on the crystal's poling design is shown in Fig. \ref{fig:JSI}a). As seen from the graphs, the experimentally measured JSI is in good agreement with the original designed one.

To quantify the symmetry of the JSA, we used the overlap integral between $f(\omega_{s},\omega_{i})$ and $f(\omega_{i},\omega_{s})$:
\begin{equation}\label{eq:overlap_int}
\eta = \left|\iint f(\omega_{s},\omega_{i})\ f^{*}(\omega_{i},\omega_{s})d\omega_sd\omega_i\right|^2,
\end{equation}
where $f=\sqrt{\frac{\text{JSI}}{\iint \mathrm{JSI} \ d\omega_sd\omega_i}}$ with JSI being the experimentally measured function. The overlap integral takes values between 0 and 1, maximized when the JSA is perfectly symmetric or antisymmetric around the degenerate line where $\omega_{s}=\omega_{i}$. The values of this integral for the theoretical and measured JSIs are 0.998 and 0.992, respectively.

The overlap integral serves as an upper limit for the degree of polarization entanglement. Denoting the lower-right lobe (see Fig. \ref{fig:JSI}) $f_{1}$ and the upper-left lobe $f_{2}$, the state vector can be written as follows:

\begin{equation}
    |\Psi\rangle=\iint \Big(f_{1}(\omega_{s},\omega_{i})a^{\dagger}_{H}(\omega_{s})a^{\dagger}_{V}(\omega_{i})+f_{2}(\omega_{s},\omega_{i})a^{\dagger}_{H}(\omega_{s})a^{\dagger}_{V}(\omega_{i})\Big)|0\rangle d\omega_{s}d\omega_{i}.
\end{equation}

The ADF transmits frequencies below the cut-off ($\omega_1<\omega_{cut}$) and reflects frequencies above the cut-off ($\omega_2>\omega_{cut}$) such that:
\begin{equation}
    a^{\dagger}(\omega_{s,i}) \rightarrow
    \begin{cases}
        b^{\dagger}(\omega_{1}) & \text{if } \omega_{s,i} < \omega_{\text{cut}} \\
        c^{\dagger}(\omega_{2}) & \text{if } \omega_{s,i} > \omega_{\text{cut}}
    \end{cases}.
\end{equation}
Following the action of the ADF on the state and the subsequent trace over the frequency, as detailed in Appendix A, the polarization density matrix can be written in the following form:
\begin{equation}
\rho_{\text{HV}}=
\begin{blockarray}{ccccc}
    \matindex{HH} & \matindex{HV} & \matindex{VH} & \matindex{VV} & \\
    \begin{block}{(cccc)c}
      0 & 0 & 0 & 0 & \matindex{HH} \\
      0 &f_{11} & f_{21} & 0 & \matindex{HV} \\
      0 & f_{12} & f_{22} & 0 & \matindex{VH} \\
      0 & 0 & 0 & 0 & \matindex{VV} \\
    \end{block}
  \end{blockarray}
\end{equation}
where the matrix elements are defined by Eq. \ref{eq:rho} as $f_{mn} = \iint f_{m}f^{*}_{n}d\omega_{1} d\omega_{2}~(m,n=\{1,2\})$ (see Appendix A). Here, the overlap between the two lobes appears explicitly as the off-diagonal terms in the two-photon density matrix. The values calculated from the measured JSI are: $f_{11} = 0.4971\pm 0.0013$, $f_{22}=0.5028 \pm0.0013$, $f_{12}=f_{21} = - 0.4978 \pm 0.0002$. We can calculate a bound on the expected quality of the polarization-entangled state from these values, which yield concurrence of $0.9956\pm0.0004$ and purity of $0.9955\pm0.0019$.

Additionally, a Schmidt analysis \cite{law2000continuous} of the JSA that corresponds to Fig. \ref{fig:JSI}(a), reveals a theoretical single-photon spectral purity $\operatorname{Tr}_{\omega_i}(\rho_{JSA}^2)$ of the total two-photon state as high as 0.496, indicating that the spectrum consists primarily of two modes.
In our case, matching the bandwidth of a pulsed femtosecond pump results in a round shape for both lobes of the JSI (see Fig. \ref{fig:JSI}(b)). Correspondingly, the value of the measured single photon spectral purity was 0.494, under the assumption of $|f|\propto \sqrt{\text{JSI}}$, with a $\pi$-phase between the lobes, which was added manually. This high single-lobe spectral purity enables the use of the presented source for protocols involving entanglement swapping, where high-visibility interference between polarization-entangled photon pairs emitted from two independent sources requires that the states from each source are pure in all degrees of freedom except polarization \cite{humble2008effects}.
Moreover, utilizing the group velocity matching (GVM) point -- which approximately holds for PPKTP at a pump wavelength of 
$\approx780$~nm and biphoton wavelengths of $\approx1560$~nm -- the phase-matching function is positioned anti-diagonally, while the pump function is aligned diagonally in $\{\omega_s,\omega_i\}$ space \cite{jin2013widely}. In this case, the source is relatively insensitive to fluctuations in the pump wavelength or in the crystal temperature.

\subsection{Certification of Polarization Entanglement}
In addition to the characterization of polarization entanglement in the spectral domain, as outlined above, we characterized the degree of polarization entanglement in the standard method of measuring the correlations between signal and idler photons in different polarization states. For this measurement, the output of the source was connected to the quantum state tomography (QST) setup \cite{White2001}  (see Fig. \ref{fig:setup}).

In the QST setup, the SMF-coupled biphoton field propagates through the ADF, after which the polarization-entangled state generated from the crystal in a single collinear beam is split into two arms of the ADF. This allows independent projections of each photon of the pair onto different polarization states, as required for QST. After the ADF, each beam is collimated using lenses $L_{3}$ and $L_{4}$, passes through a quarter-wave plate (QWP$_{2(3)}$) and a polarizer (Pol), and is then focused by lenses $L_{5}$ and $L_{6}$ onto two superconducting nanowire detectors ($\mathrm{Det}_{1,2}$) \cite{scontelSNSPD}. The pump power was 50 mW.

\textcolor{black}{The two output fibers of the ADF introduced a phase difference between the H and V polarizations due to twist-induced birefringence, which differed between the fibers. As a result, the original $\pi$-phase difference between the two terms (lobes) of the generated state was not preserved. A tiltable quarter-wave plate (QWP$_{1}$) placed in one of the arms after the ADF was used to fine-tune the phase between H and V polarizations. Adjusting this phase allowed us to set the relative phase between the two terms of the state to $\pi$, leading to the preparation of the $|\Psi^{-}\rangle$ Bell state.}

\begin{figure}[t]
  \centering
  \includegraphics[width=\textwidth]{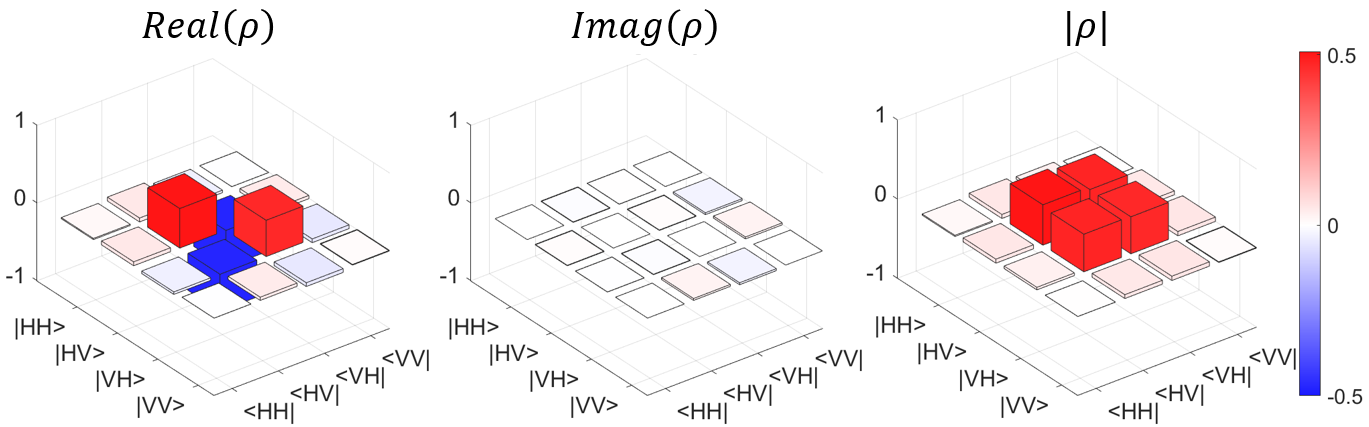}
\caption{Reconstructed density matrix $\rho$. From left to right: real, imaginary, and absolute values of $\rho$ are shown. Based on this matrix, the purity is $P=0.948\pm0.003$, the concurrence is $C=0.948\pm0.004$, and the fidelity with respect to the $|\Psi^{-}\rangle$ state is $0.963\pm0.001$. The CHSH parameter is $S=2.747\pm0.004$.}
\label{fig: density matrix}
\end{figure}

Following the 16-measurement QST scheme \cite{White2001}, the maximum-likelihood density matrix of the generated biphoton state was experimentally reconstructed (Fig. \ref{fig: density matrix}). 
Since the density matrix fully describes the generated polarization state, all other parameters characterizing the quantum state can be derived from it. Specifically, based on the measured density matrix, we calculated the purity $p=\text{Tr}(\rho^2)$ to be $0.948\pm0.003$, and the concurrence \cite{hill1997entanglement,hildebrand2007concurrence,RevModPhys.81.865,wootters1998entanglement} to be $0.948\pm0.004$. For Bell-type tests, the generated state violates the Clauser-Horne-Shimony-Holt (CHSH) inequality \cite{clauser1969proposed} with an $S$-parameter of $S=2.747\pm0.004$.  


\textcolor{black}{Based on the measured density matrix, the Bell fidelity $F$ with respect to the $|\Psi^{-}\rangle$ state was calculated to be $0.963\pm0.001$. As a result, the purity and concurrence of the generated state were characterized via both JSI measurements and QST.} Although the spectral method provides insight into the state's frequency correlations and expected purity, it does not directly characterize polarization entanglement. The polarization-based QST method, on the other hand, reconstructs the full density matrix but is more sensitive to experimental imperfections, such as wave plate misalignment, non-collinear generation, and the finite extinction ratio of the add-drop filter. Additionally, the entanglement visibility measured in the {D,A} basis reflects the degree of quantum interference. It is directly influenced by multi-photon events, which become more prominent at higher pump powers. Therefore, the spectral and polarization methods serve complementary roles in assessing state quality, with the spectral method predicting purity from energy correlations and QST providing a direct measurement of polarization entanglement.

\subsection{Visibility measurements and the effect of high pump power}

\begin{figure}[t]
  \centering
  \includegraphics[width=1\textwidth]{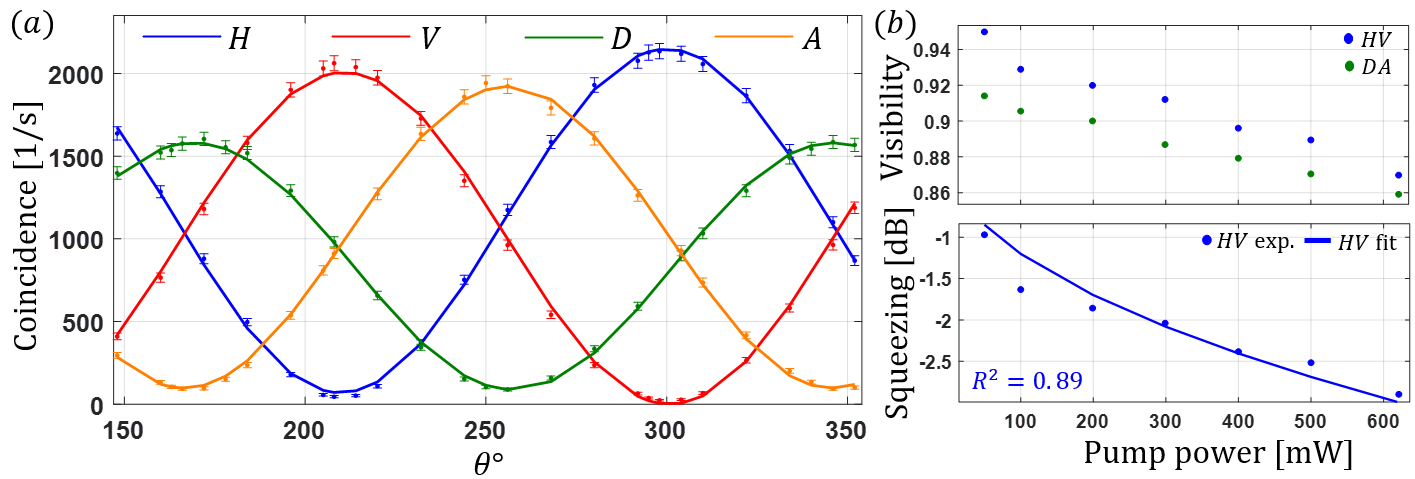}
\caption{(a) Polarization entanglement visibility. The measure data points are fitted  with sine function: $a\sin^{2}(b\theta+c)+d$, then $\mathcal{V}=a/(a+2d)$. $\mathcal{V}_{H} = 0.94$, $\mathcal{V}_{V} = 1.00$, $\mathcal{V}_{D} = 0.90$, $\mathcal{V}_{A} = 0.90$.  $R$-square > 0.99 (for all four fits). Pump power $\approx 45$ mW. (b) The polarization entanglement visibility (top) and the squeezing (bottom) in \{H,V\} and \{D,A\} bases as a function of pump power. The \{H,V\} basis set of visibility measurements was used to calculate the squeezing. The squeezing as a function of pump power was then fitted to $r = C\sqrt{P_p}$, where $P_p$ is the pump power and $C$ is a constant.
}
\label{fig:vis}
\end{figure}

We also performed polarization visibility measurements \cite{White2001,altepeter2005photonic}. The QWP$_{2}$ and QWP$_{3}$ in both arms of the QST section were removed for these measurements. The results are shown in Fig. \ref{fig:vis}a. Each of the curves was measured by setting the polarizer in one arm at $0^{\circ},90^{\circ},45^{\circ}$, and $-45^{\circ}$ in the laboratory reference frame, corresponding to the $H, V, D$, and $A$ curves in Fig. \ref{fig:vis}a, respectively, while rotating the other polarizer. The measurement data points were fitted with sine functions. From the fitting curves, the average visibility in the \{H,V\} basis was 97\%, and 90\% in the \{D,A\} basis at a pump power of 45 mW. 

Since photon pairs are generated in the crystal in the \{H,V\} basis, the visibility measured in this basis remains high even when the photons are not fully polarization-entangled. In contrast, the visibility on the \{A,D\} basis directly reflects the degree of quantum interference and entanglement. Due to imperfections in the crystal design and limitations in the experimentally achievable state fidelity - such as the collection of multiple spatial Schmidt modes, non-collinearly generated photons, and the finite extinction ratio of the add-drop filter - the visibility in the A/D basis is lower than in the H/V basis, as typically observed in other types of sources \cite{PhysRevA.73.012316,Cai:22,predojevic2012pulsed}.

By directly connecting the two output fibers of the ADF to the detectors and measuring a coincidence rate of $\mathcal{R}_{c}\approx6$~kHz, along with singles rates of $\mathcal{R}_{s}\approx\mathcal{R}_{i}\approx20$~kHz at 1 mW of pump power, we determined the source's generation rate to be $\mathcal{R}_{gen} = \mathcal{R}_{s}\mathcal{R}_{i}/\mathcal{R}_{c} \approx67$ kHz/mW with an average heralding efficiency of $\eta_{H}=\mathcal{R}_{c}/\sqrt{\mathcal{R}_{s}\mathcal{R}_{i}} \approx 0.3$.

The field generated in the SPDC process is a coherent superposition of two-mode Fock states correlated by photon number (two-mode squeezed vacuum) \cite{schneeloch2019introduction}. Thus, the biphoton state represented by Eq. \ref{eq:initial state} is only one term in the complete superposition, with the remaining terms responsible for multi-pair generation events (the vacuum state term is eliminated by the coincidence measurement of the two detectors \cite{Yesharim_review_2025}).
These multi-photon events cause the generated state to deviate from the ideal two-photon state represented by Eq. \ref{eq:initial state}, gradually reducing the quality of entanglement as the probability of SPDC increases. A higher pump power is a key contributing factor to this increased probability of SPDC.
\textcolor{black}{The theoretically calculated  coupling coefficient, $\kappa$ \cite{hurvitz2023}, was $\kappa = 0.5$ and $\kappa = 1.8$ for pump powers of $50\mathrm{mW}$ and $620\mathrm{mW}$, respectively. This indicates that the system operates in the high-gain regime at high pump powers.}
To analyze the effect of pump power on our source, we measured the visibility in the \{H,V\} and \{D,A\} bases as a function of the pump power. The results are demonstrated in Fig. \ref{fig:vis}(b) top. The graph demonstrates that even at relatively high pump powers, the visibility remains well above the classical limit of 71\% \cite{MOTAZEDIFARD2021e07384}. 

The measured reduction in visibility at high pump powers is due to the generation of high-order terms of the SPDC Hamiltonian \cite{Takesue2010}. This reduction enables the estimation of the mean photon number $\mu = \sinh^2(r)$ and the corresponding squeezing parameter, $S = e^{-2r}$, as shown in Fig. \ref{fig:vis}(b) bottom. These measurements are in reasonable agreement with a simple fit of $r$ to the pump power, $r=C\sqrt{P_{pump}}$. For the highest pump power of $620$mW, we estimate from the visibility reduction that $\mu=0.1$ with squeezing of $-3 \mathrm{dB}$ for the \{H,V\} configuration. At this average pump power, the peak power is $80\mathrm{kW}$ and the peak intensity is $0.9\mathrm{GW/cm^2}$.

We analyzed the squeezing in the \{H,V\} basis as in principle, the amount of squeezing should be the same across mutually unbiased bases. Experimental imperfections such as deviations from the designed JSI and non-ideal waveplates lead to reduced visibility in the \{D,A\} basis. These imperfections introduce distinguishability between the photon-pair components and affect the interference quality, resulting in an apparent variation in the extracted squeezing parameter across different bases. Since the \{H,V\} basis is less affected by such artifacts and provides the most reliable visibility, we consider it the most accurate basis for estimating the squeezing level.

\section{Conclusions}
We have demonstrated a simple, efficient, and versatile source of polarization-entangled photon pairs using a single-pass, domain-engineered KTP crystal.  Our method eliminates the need for additional optical components, improving stability and compactness.  High-quality entanglement was achieved (CHSH parameter S = 2.747 ± 0.004). The source will operate under both pulsed and continuous-wave pumping (provided the symmetric JSA condition is satisfied), making it suitable for various quantum applications. The high spectral purity of single photons from each lobe of the generated biphotons JSI makes it particularly promising for entanglement swapping protocols that are central to the future quantum internet. Furthermore, this method can be implemented using nonlinear optical waveguides or optical cavities, to enhance the total pair generation rate or spectral brightness.

We presented a method for the direct reconstruction of the polarization-entangled state’s density matrix from the measured JSI. Quantifying the spectral symmetry of the JSI, enables the estimation of the purity and concurrence of the polarization-entangled state directly from spectral measurements, offering a highly efficient and experimentally accessible approach. The results are in good agreement with a standard quantum state tomography method.

We have also measured the pump power dependence of the polarization entanglement visibility in the \{H,V\} and  \{D,A\} bases. Increasing the pump power leads to the generation of multiple signal-idler pairs, and reducing the observed visibility. Thus, we estimate a squeezing level of approximately $-3 \mathrm{dB}$ at high pump power of $620$ mW. All of the above measurements and the derived parameters demonstrate the quantum nature of the generated state.

Whereas we have described the frequency degree of freedom using discrete frequency bins, we can extend the analysis to consider the frequency (and hence time) as continuous variables \cite{PhysRevA.111.L030403}. Furthermore, here we considered round JSA lobes, but other shapes (such as elliptical lobes) can be obtained by shaping the pump spectrum and phase matching functions \cite{Shukhin2024}, thus opening new possibilities for the generation of time-energy entangled states.

\section*{Appendix A: Derivation of the polarization density matrix}

Then the action of the ADF on the incident quantum state is:
\begin{equation}
\begin{split}
|\Psi\rangle_{\text{ADF}}&= \iint \Big(f_{1}(\omega_{1},\omega_{2})b^{\dagger}_{H}(\omega_{1})c^{\dagger}_{V}(\omega_{2})+f_{2}(\omega_{2},\omega_{1})c^{\dagger}_{H}(\omega_{2})b^{\dagger}_{V}(\omega_{1})\Big)|0\rangle d\omega_{1}d\omega_{2} \\
&= \iint \Big(f_{1}(\omega_{1},\omega_{2})|\omega_{1_{H}}\rangle |\omega_{2_{V}}\rangle + f_{2}(\omega_{2},\omega_{1})|\omega_{2_{H}}\rangle |\omega_{1_{V}}\rangle \Big) d\omega_{1}d\omega_{2}.
\end{split}
\end{equation}
The density matrix is then written as follows:
\begin{equation}
\begin{split}
\rho & = |\Psi\rangle_{\text{ADF}} \langle \Psi |_{\text{ADF}} \\
& = 
\iiiint d\omega_{1}d\omega_{2}d\omega'_{1}d\omega'_{2}\Big[ \Big( f_{1}(\omega_{1},\omega_{2})|\omega_{1_{H}}\rangle |\omega_{2_{V}}\rangle + f_{2}(\omega_{2},\omega_{1})|\omega_{2_{H}}\rangle |\omega_{1_{V}}\rangle \Big) \times \\
& \times \Big( f_{1}^{*}(\omega'_{1},\omega'_{2}) \langle\omega'_{1_{H}}| \langle\omega'_{2_{V}}| + f_{2}^{*}(\omega'_{2},\omega'_{1}) \langle\omega'_{2_{H}}| \langle\omega'_{1_{V}}|\Big)\Big] \\
&=\iint d\omega_{1}d\omega_{2} \Big( 
f_{1}(\omega_{1},\omega_{2})f^{*}_{1}(\omega_{1},\omega_{2})
|\omega_{1_{H}}\rangle|\omega_{2_{V}}\rangle\langle\omega_{1_{H}}|
\langle\omega_{2_{V}}| + \\
& + f_{1}(\omega_{1},\omega_{2})f^{*}_{2}(\omega_{2},\omega_{1})
|\omega_{1_{H}}\rangle|\omega_{2_{V}}\rangle\langle\omega_{2_{H}}|
\langle\omega_{1_{V}}| + \\
& + f_{2}(\omega_{2},\omega_{1})f^{*}_{1}(\omega_{1},\omega_{2})
|\omega_{2_{H}}\rangle|\omega_{1_{V}}\rangle\langle\omega_{1_{H}}|
\langle\omega_{2_{V}}| + \\
& + f_{2}(\omega_{2},\omega_{1})f^{*}_{2}(\omega_{2},\omega_{1})
|\omega_{2_{H}}\rangle|\omega_{1_{V}}\rangle\langle\omega_{2_{H}}|
\langle\omega_{1_{V}}|\Big),
\end{split}
\end{equation}
where $\omega_{1} = \omega'_{1}$ and $\omega_{2} = \omega'_{2}$. After rewriting $|\omega_{1_{H}}\rangle\langle\omega_{1_{H}}|$ as $|\omega_{1}\rangle|H\rangle\langle\omega_{1}|\langle H|$ (and similar expressions for all other terms) and tracing out the frequency states, the density matrix in the polarization basis is written as follows:

\begin{equation}\label{eq:rho}
\begin{split}
    \rho_{\text{HV}} &= \iint d\omega_{1}d\omega_{2} 
    \Big( |f_{1}(\omega_{1},\omega_{2})|^{2} |H\rangle|V\rangle\langle H| \langle V| + \\
    &+f_{1}(\omega_{1},\omega_{2})f^{*}_{2}(\omega_{2},\omega_{1})  |H\rangle|V\rangle\langle V| \langle H|+f_{2}(\omega_{2},\omega_{1})f^{*}_{1}(\omega_{1},\omega_{2})  |V\rangle|H\rangle\langle H| \langle V| +\\
    &+|f_{2}(\omega_{2},\omega_{1})|^{2} |V\rangle|H\rangle\langle V| \langle H|
    \Big).
\end{split}
\end{equation}

\begin{backmatter}
\bmsection{Funding}
Israel Science Foundation, grants 969/22 and 3117/23.

\bmsection{Acknowledgments}
We thank Ziv Gefen from Raicol Crystals for his assistance in preparing the nonlinear crystals.
I.H. acknowledges the Weinstein Research Institute for signal processing support and the Israeli planning and budgeting committee for quantum science and technology research.

\bmsection{Disclosures}
The authors declare no conflicts of interest.


\bmsection{Data availability} Data underlying the results presented in this paper are not publicly available at this time but may be obtained from the authors upon reasonable request.



\end{backmatter}


\bibliography{sample}

\end{document}